\documentclass[]{aa}

\usepackage{graphicx,txfonts}

\newcommand{\oasis}{{\tt OASIS}}
\newcommand{\sauron}{{\tt SAURON}}
\newcommand{\stis}{{\tt STIS}}
\newcommand{\pueo}{{\tt PUEO}}

\newcommand{\ten}[1] {10$^{#1}$}

\newcommand{\kms}{km.s$^{-1}$}
\newcommand{\etal}{et al.}

\newcommand{\eg}{\emph{e.g. }}

\begin{document}

\title{A 60~pc counter-rotating core in NGC~4621\thanks{Based on observations collected at Canada-France-Hawaii Telescope, which is operated by CNRS of France, NRC of Canada, and the University of Hawaii.}}

\author{Fabien Wernli \inst{1}, Eric Emsellem \inst{1}, Yannick Copin \inst{2}}

\offprints{F. Wernli,~\email{fabien@obs.univ-lyon1.fr}}

\institute{Centre de Recherche Astronomique de Lyon, 9 av. Charles Andr\'e,
  69561 Saint-Genis Laval Cedex, France \and
  Institut de physique nucl\'eaire de Lyon, 4 rue Enrico Fermi, 69622 Villeurbanne Cedex, France}

\date{Accepted September 10 2002}

\abstract{ We present adaptive optics assisted \oasis\ integral field
  spectrography of the S0 galaxy NGC~4621. Two-dimensional stellar kinematical
  maps (mean velocity and dispersion) reveal the presence of a $\sim 60$~pc
  diameter counter-rotating core (CRC), the smallest observed to date. The
  \oasis\ data also suggests that the kinematic center of the CRC is slightly
  offset from the center of the outer isophotes. This seems to be confirmed by
  archival HST/\stis\ data. We also present the HST/WFPC2 $V-I$ colour map,
  which exhibits a central elongated red structure, also slightly off-centered
  in the same direction as the kinematic centre. We then construct an
  axisymmetric model of NGC~4621: the two-integral distribution function is
  derived using the Multi-Gaussian Expansion and the Hunter \&
  Qian~(\cite{hun}) formalisms.  Although the stellar velocities are
  reasonably fitted, including the region of the counter-rotating core,
  significant discrepancies between the model and the observations demonstrate
  the need for a more general model (\eg a three-integral model).


}
\authorrunning{F. Wernli \etal}
\maketitle


\section{Introduction}

The fact that early-type galaxies are not necessarily oblate major-axis
rotators has been well known since the end of the 1970s. The number of objects
showing significant minor-axis rotation attained ten in 1988 (Wagner
\etal~\cite{wag88}), and some of these objects later showed evidence for
kinematically decoupled core structures (NGC~4365, NGC~4406:
Bender~\cite{ben88}; NGC~4589: Moellenhoff \& Bender~\cite{moe89}; NGC~5982:
Wagner~\cite{wag90}).  The most cited formation scenario for such structures
is the hierarchical model, which involves single or multiple merging steps
(see \eg \cite{hierarch_col} and references therein).  Observations of
kinematically decoupled cores can therefore be used to constrain the merger
tree of galaxies.  However the total fraction of early-type galaxies
containing such a core is poorly known (de Zeeuw \& Franx~\cite{dez91}).  Such
cores are generally revealed using long-slit measurements with 1\arcsec\ FWHM
or worse spatial resolution: this introduces an a priori assumption on the
geometry of the central structures, and prevents the detection of cores with
small apparent sizes.  Finally, the rough characteristics of these cores
(fraction in mass, relative age and metallicity) remain very uncertain, as
very few studies exist and are often limited to morphological data (see
Carollo \etal~\cite{marc} for 15 KDCs).

Recent studies (\eg Verolme \etal~\cite{ellen}) show that state-of-the-art
models combined with two-dimensional integral-field spectroscopic data are
required to constrain precisely the global physical parameters of early-type
galaxies, such as the inclination angle or the mass-to-light ratio, as well as
the central characteristics for objects exhibiting features such as
kinematically decoupled cores (\eg the core mass of IC~1459, Cappellari \etal~
\cite{ic1459}). Moreover, integral-field spectroscopy also provides access to
line-strength maps: the detailed kinematic information can thus be coupled to
the line indices in order to more easily disentangle the core from the host
galaxy (Davies et al.  2001).

The \sauron\ survey (Bacon \etal~\cite{paperI}, de Zeeuw \etal~
\cite{paperII}) was carried out in order to study the integral field
kinematics of a sample of 72 early-type galaxies over a wide field of view ($>$
33\arcsec$\times$41\arcsec). In parallel, to obtain complementary high spatial
resolution data, we designed a program to observe their central parts
($\sim$\mbox{5\arcsec$\times$5\arcsec}) with the integral field spectrograph
\oasis\ (Bacon \etal~\cite{oasis}). The aim is to probe the central spatial
morphology and dynamics of a subset of galaxies in the \sauron\ sample, in
order to link them with the global properties of the host galaxies provided by
\sauron\ and other wide field studies. The additional use of {\tt PUEO}
(Rigaut~\etal~\cite{pueo}), the Canada France Hawaii Telescope AO bonnette,
with \oasis\ allows us to reach resolutions of about $0\farcs25$ ($\sim~12$~pc
at 10~Mpc) in the red part of the visible when bright guiding sources are
available.

NGC~4621, an S0 galaxy (Lauer \etal~\cite{paper1}), with $M_V=9.6$, located in
the Virgo cluster (D=18.3 Mpc, Tonry \etal~\cite{sbf_ton}), is the first
target of the subsample that was observed with \oasis/\pueo.  Its steep
power-law central luminosity profile ($\gamma = 2.03$,
Gebhardt~\etal~\cite{paper3}) is ideally adapted to serve as a guiding source
for the AO.  NGC~4621 is close to edge-on, and its innermost $6\arcsec\ $
isodensities reveal a nuclear disk (Sil'Chenko~\etal\ \cite{silchenko}).

In this paper, we present the first results of our \oasis\ observations of
NGC~4621, with the discovery of the smallest counter-rotating stellar core
observed to date. The scale of the core is 60~pc, while the previously
detected cores have an average radius of around 1~kpc (see the sample of
Carollo \etal~\cite{marc}). Detailed information on the different data sets
used in this work, including HST/WFPC2 and \stis\ observations, are provided
in Sect.~\ref{sec:data}.  The corresponding measured photometry and stellar
kinematics are presented in Sect.~\ref{sec:results}. We computed a
two-integral distribution function (DF) model of the galaxy, which is
presented in Sect.~\ref{sec:DF}. The discussion is carried out in
Sect.~\ref{sec:discussion}.


\section{Observational data sets}
\label{sec:data}

\subsection{\oasis\ data}

\subsubsection{Observations}

Ca-triplet \oasis\ exposures of NGC~4621 were obtained in January 2000 using
the medium spectral configuration (MR3: 8346-9152~\AA) at the f/20 ({\tt
  PUEO}) focus of the CFHT, and a spatial sampling of 0\farcs16. Eight out of
ten 30~mn exposures were fully reduced using the dedicated {\tt XOasis}
package\footnote{{\tt http://www-obs.univ-lyon1.fr/\~{}oasis/home/index.html}},
and merged into a single frame (two frames were discarded due to bad seeing
conditions and associated loss of guiding). The resulting 3D
dataset\footnote{$(\alpha,\delta,\lambda)$, one spectrum per spatial element}
probes about $4\arcsec\times4\arcsec$ (300$\times$300~pc$^2$), with a
resolution of $\sigma=70$~\kms. We modeled the PSF of the merged
datacube using the sum of two concentric Gaussians: their parameters were
retrieved (via a least-square procedure) by comparison with the WFPC2/HST
F814W frame as in Bacon \etal~(\cite{m31}). The resulting FWHM of the merged
datacube is 0\farcs51.

The total S/N (summed over the bandwidth) drops very quickly towards the outer
parts, as it is 65 at the center, 20 at 1\arcsec\ and 10 at 2\arcsec on the
major-axis. We thus convolved the datacube with a Gaussian of variable width in
order to obtain a higher S/N over the field (see Fig.~\ref{fig:VGH}). 

\subsubsection{Kinematic measurements}

We used a tuned version of the Fourier Correlation Quotient (FCQ;
Bender~\cite{bender}) method to derive the line of sight velocity
distributions (LOSVD) at each measured spatial position. We used a single
template star (G8III: HD073665), but tested the stability of the
kinematical mesurements with different templates, and different continuum 
subtraction parameters, only resulting in minor differences. 
The LOSVDs were then fitted using a single
Gaussian function, yielding the mean velocity and dispersion maps.
We could not obtain reasonable maps of higher order moments, due
to the rather low signal to noise ratio (even after the convolution performed 
on the datacube). The error bars on the velocity and dispersion measurements
were computed via a Monte-Carlo algorithm using the measured S/N 
(see Copin \cite{copin}, PhD Thesis).

\subsection{Ancillary data}

\subsubsection{Photometry}

In order to obtain an accurate three-dimensional mass model of the galaxy we
need both wide-field and high-resolution photometry to probe the visible mass
up to large radii and to correctly sample the central structure (the cusp).
We used a 20\arcmin\ $V$ band image of the galaxy taken at the OHP 2m
Telescope, kindly provided by R. Michard (Idiart, Michard, \& de Freitas
Pacheco \cite{mich}). The central regions were examined with the help of
WFPC2/HST data\footnote{based on observations collected with the NASA/ESA HST,
  obtained at STScI, which is operated by AURA, Inc, under NASA contract
  NAS5-26555} retrieved from the archive (F555W and F814W filters,
Faber~\etal~\cite{paper4}, \#5512). For both bands, three unsaturated frames
were merged and cosmic ray-corrected (330~s and 230~s exposure time for F555W
and F814W filters respectively). We performed photometric calibration using
the {\tt VEGAMAG} standard, and used the PSFs computed using {\tt
  tinytim\footnote{\tt http://www.stsci.edu/software/tinytim/tinytim.html}}.
We adjusted the levels of the OHP image to the F555W frame taking into account
the different spectral domains and the PSF.

\subsubsection{Long-slit kinematics}

We used long slit data from Bender, Saglia, \& Gerhard (\cite{BSG94},
hereafter BSG94; data kindly provided by R.~Bender) in order to complement the
\oasis\ data up to large radii: major and minor axis velocity and dispersion
measurements are available inside 40\arcsec\ (3.5~kpc), with a seeing of
$\sim$1\farcs8 FWHM (other kinematical - and line-strength - measurements were
published by Sil'chenko \cite{silchenko}, but with a lower spatial resolution).

We finally reduced unpublished {\tt HST/STIS} major-axis data (Green,
ID~\#8018; G750M grating, Ca-Triplet) in order to examine the central
kinematics of NGC 4621 at high spatial resolution.  Two exposures with a total
of 72~mn were available.  The calibrated data were retrieved via the STScI
data archival system ({\tt calstis} pipeline). Appropriate flat field
exposures were retrieved to correct for the fringing ({\tt
  mkfringeflat/defringe} \emph{IRAF} routines, Goudfrooij \& Christensen
\cite{defringe}), critical at these wavelengths, as well as five K0-III
kinematical template star exposures (HR7615, Green, \#7566). Further rejection
of cosmic rays was then performed on the NGC~4621 individual exposures, before
recentring and merging.  The stellar kinematics were extracted using the same
FCQ and fitting methods as for the \oasis\ data.  As for the \oasis\ 
datacubes, the S/N of the {\tt STIS} data was found sufficient only to derive
the mean stellar velocity and dispersion within the central arcsecond.


\section{Results}
\label{sec:results}

To avoid any confusion due to the complexity in both the photometry and the
kinematics of NGC~4621, we will use the following convention: the photometric
major axis of the galaxy (as measured by HST) defines the $x$-axis, the
negative values being in the SE quadrant, referred to as the left side of the
galaxy. The center $(0,0)$ is defined as the center of the HST isophotes of
NGC~4621 within 1\arcsec, excluding the central 0\farcs1, the isophotes of
which exhibit a significant asymmetry (see next Section). The centering and
rotation procedures of the HST frames have been performed using an algorithm
which minimizes the standard deviation of the recentered and rotated frame
subtracted by its flipped counterpart. This leads to accuracies of 0.1 degrees
and 0\farcs005. All figures in this paper share a common orientation, as shown
in Fig.~\ref{fig:center}.

\subsection{Stellar velocity and velocity dispersion}
\label{sec:kin}

\begin{figure*}[tbp]
  \centering
  \resizebox{\textwidth}{!}{\includegraphics{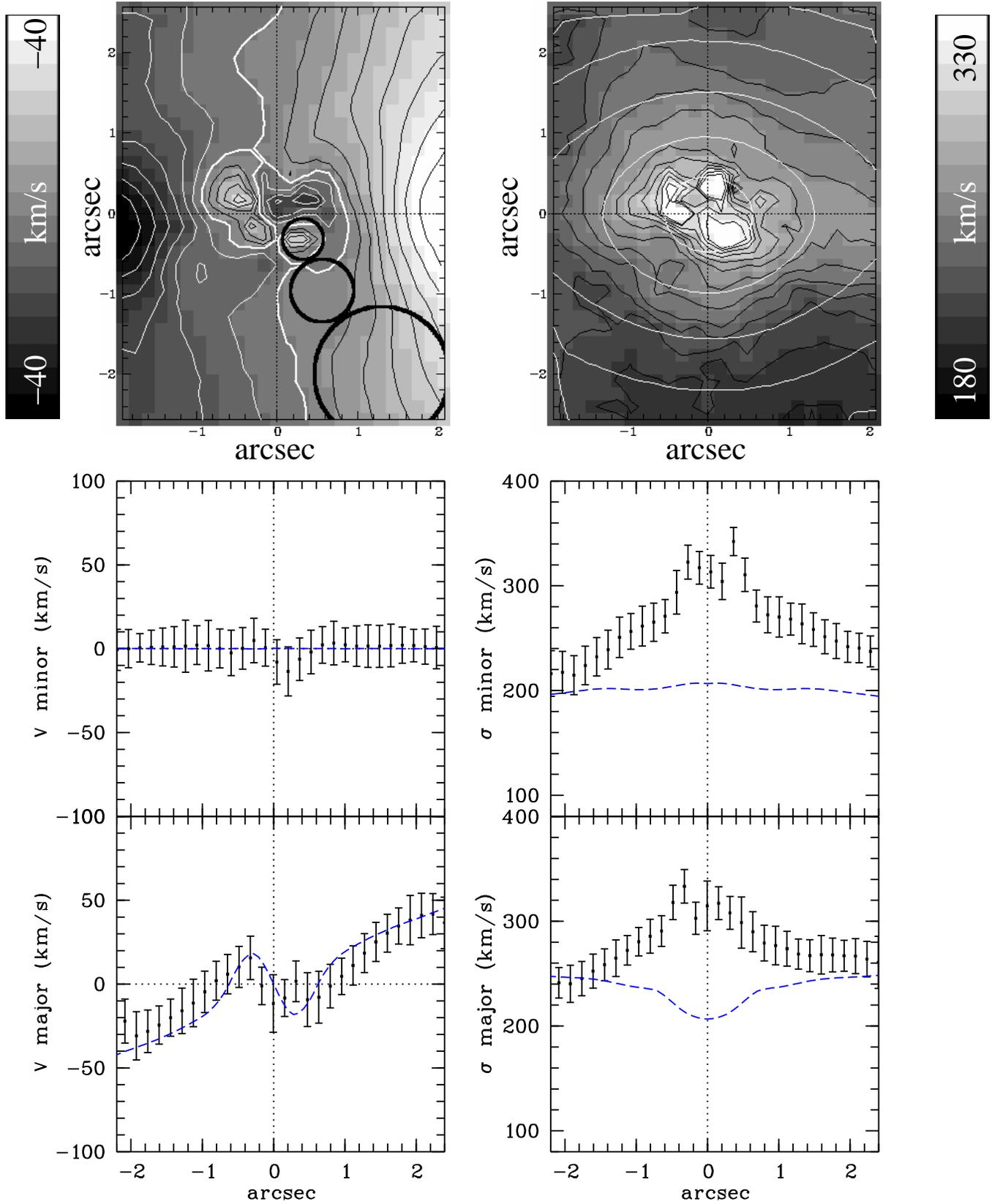}}
    \caption{\oasis\ stellar kinematic data of NGC 4621. Top left: mean
      velocity field. The zero-velocity curve has been emphasized with a white
      thick line.  Top right: velocity dispersion. White contours are
      reconstructed isophotes. The data have been smoothed to improve the
      signal to noise: the black circles represent the beam sizes (FWHM).
      Below the cuts along the major and the minor axes are shown along with
      the best fitting two-integral model (dashed lines, see
      Sect.~\ref{sec:DF}).}
    \label{fig:VGH}
\end{figure*}

The \oasis\ stellar mean velocity and dispersion maps are presented in
Fig.~\ref{fig:VGH}. The datacubes have been centered and rotated in order to
match the HST data, using the center and angle provided by the PSF fitting
procedure mentioned in Section \ref{sec:data}. The velocity field reveals a
clear counter-rotating core (CRC). The position of the zero-velocity curve is
used to measure the center and the size\footnote{The limits of the CRC are
  defined by the zero-velocity curve} of the CRC: it is 1\farcs7 (150~pc) in
diameter, and off-centered by $\sim 0\farcs2$ towards the SE.  The total
velocity amplitude observed within the central 1\arcsec\ along the major axis
is 100~\kms, while the peak-to-peak velocity amplitude of the CRC's is only
35~\kms. The stellar velocity dispersion map peaks at $\sim 330 \pm 25$~\kms.
The dispersion map however exhibits high frequency substructures in the
central 0\farcs5 (Fig.\ref{fig:VGH}), the minor-axis dispersion profile even
having a local minimum at the centre.  Considering the amplitude of these
structures and their spatial scales (comparable to the local FWHM), we should
wait for data with better signal to noise ratio to discuss these features.

The BSG94 data show no significant minor-axis rotation, and a maximum velocity
of $140$~\kms\ at 30\arcsec (see BSG94 or Fig.~\ref{fig:plotcompn}).  Besides
a relatively weak velocity amplitude in the innermost 4\arcsec\ ($<$\ 
60~\kms), these data do not show any hint for the existence of the
counter-rotating core. We will use the BSG94 kinematics to constrain the
dynamical model outside the \oasis\ field of view.

\begin{figure*}
  \begin{center}
    \begin{tabular}{cc}
      \includegraphics{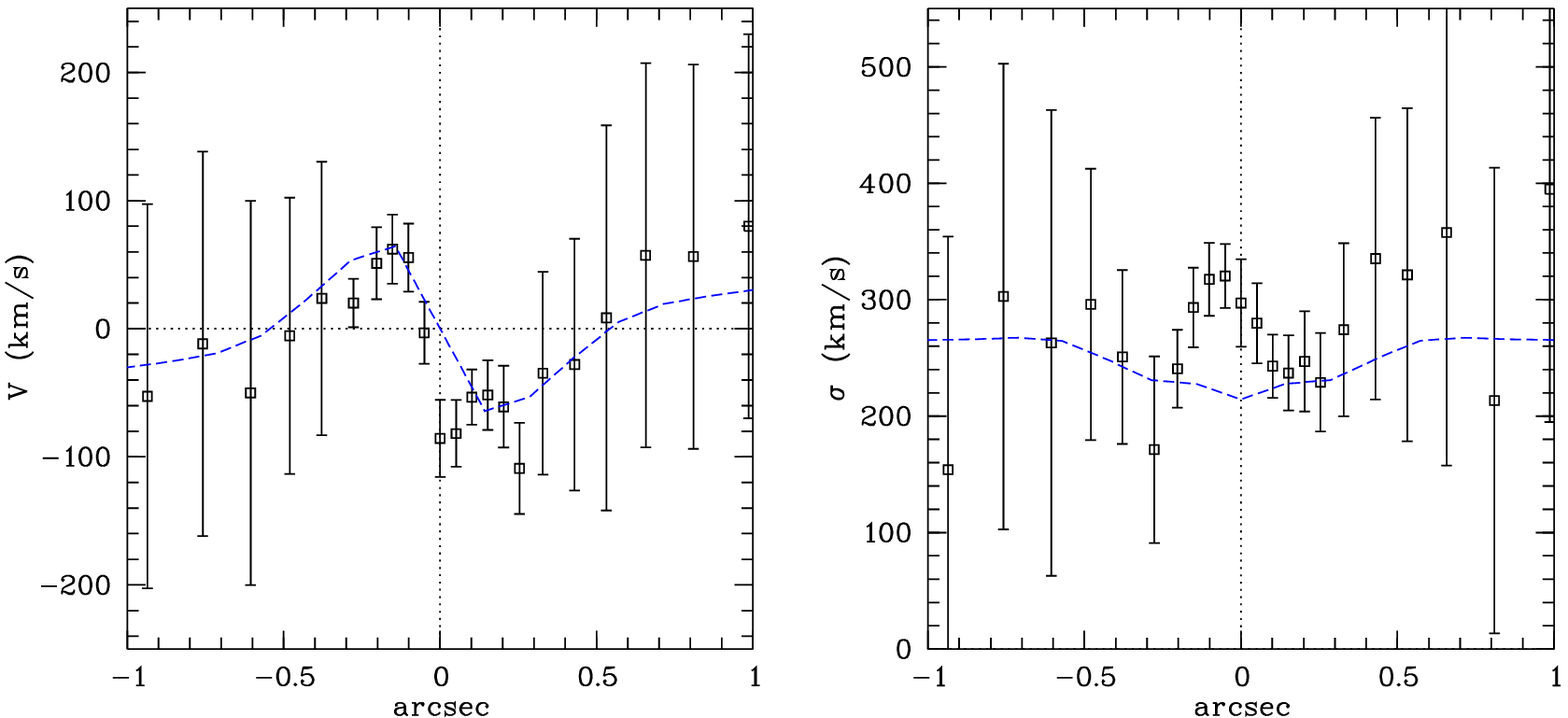}
    \end{tabular}
    \caption{\stis\ mean velocity (left) and dispersion (right) profiles along
      the major axis. The dashed line corresponds to the best fitting
      two-integral model (see Sect.~\ref{sec:DF}). The velocity dispersion is
         clearly underestimated by the present model, which hints for the need
         of an additional central dark mass, a more general treatment
         (three-integral model) or both.}
    \label{fig:HRdata}
  \end{center}
\end{figure*}

The \stis\ velocities and dispersions are shown in Fig.~\ref{fig:HRdata}. In
these data the size of the CRC is 0\farcs7 ($\sim 60$~pc) and its velocity
center is located at $x=-0\farcs05$ ($\sim 4.5$~pc). We have carefully
checked the centering of the \stis\ data with respect to 
the reference WFPC2 images, and found that this offset is robust.
At \stis\ resolution the major-axis velocities of the CRC reach
$\sim 75$~\kms with respect to the systemic. The maximum velocity 
dispersion is $320\pm27$~\kms, at $x = 0\farcs05$.  
These values are consistent with the \oasis\ data considering the
higher spatial resolution of the \stis\ data.

\subsection{HST photometry}

\begin{figure}[h]
  \begin{center}
    \resizebox{7.5cm}{!}{\includegraphics{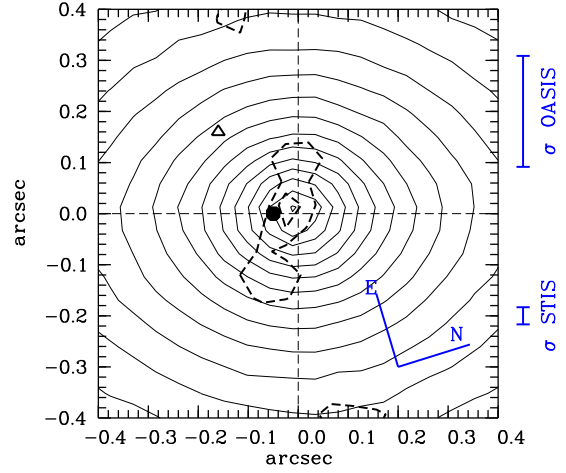}}
    \caption{Inner isophotes of NGC~4621 (WFPC2 F555W, solid contours, step of
      0.2~mag). The center (0,0) of the galaxy is defined as the center of the
      outer isophotes (outside 0\farcs1). The open triangle and filled circle
      correspond to the center of the counter-rotating core (CRC) as measured
      in the \oasis\ and \stis\ data repectively (see Sect.~\ref{sec:kin}).
      The vertical bars correspond to the \oasis\ and \stis\ spatial
      resolutions ($\sigma$). The dashed contours represent the $V-I$ colour
      map from WFPC2: levels 1.36 and 1.37 mag. North is 73 degrees
      clockwise from the vertical axis.}
    \label{fig:center}
  \end{center}
\end{figure}

Both F555W and F814W images reveal a peculiar structure in the core. Indeed,
the luminosity peak is offset from the center of the outer isophotes by
0\farcs01 towards the east. The photometric peak is located in the upper
left quadrant (Fig.~\ref{fig:center}). This feature does not seem to be an
artefact, as it is clearly visible on each individual HST frame, and in both
bands, and cannot be attributed to centering uncertainties (0\farcs005).

\begin{figure*}
  \begin{center}
    \resizebox{\textwidth}{!}{\includegraphics{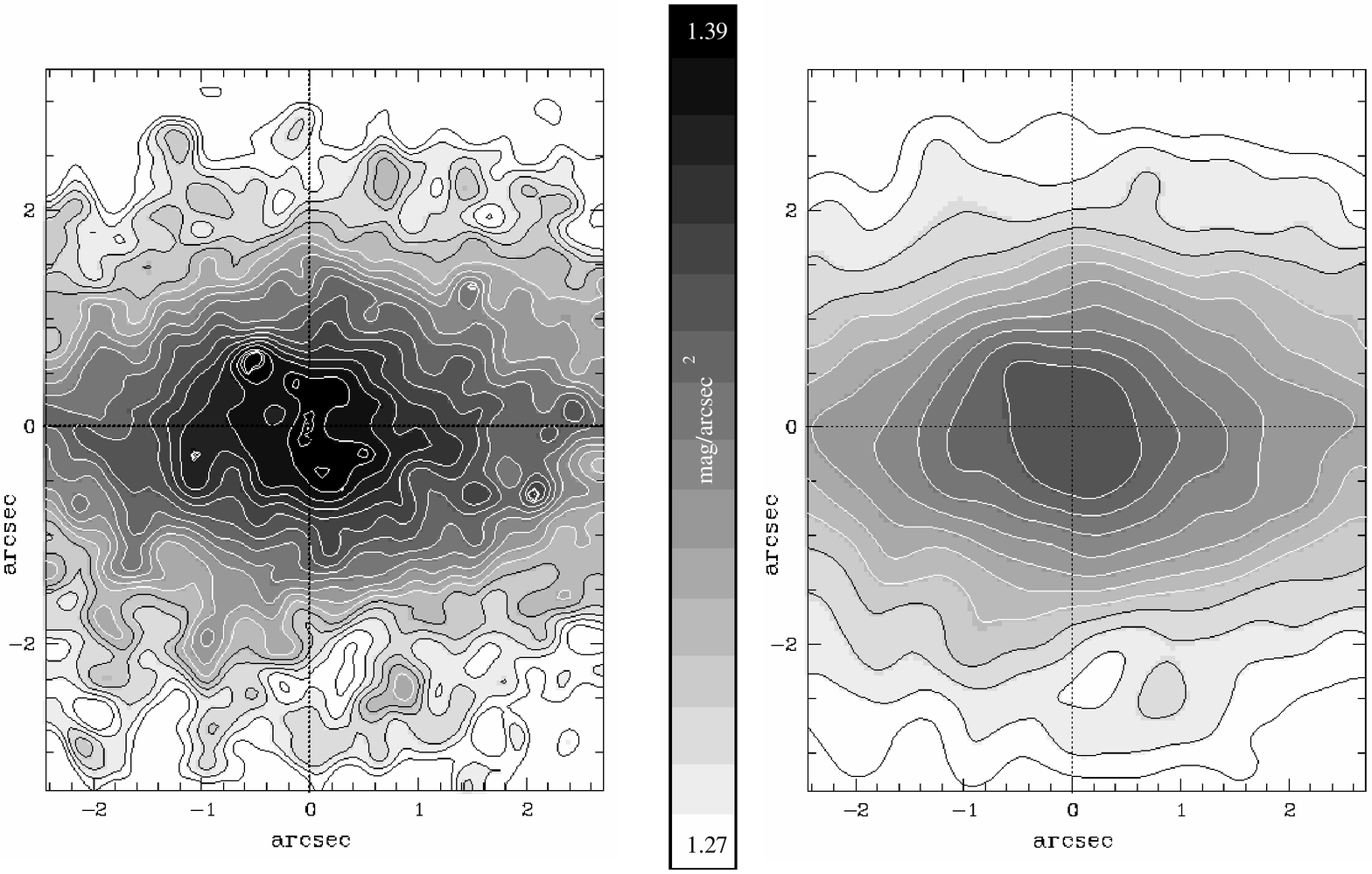}}
    \caption{$V-I$ map (WFPC2, F555W and F814W filters). Left: convolved with
      a gaussian of $\sigma=0\farcs1$. Right: convolved with the PSF of the
      \oasis\ data (0\farcs51 FWHM). Contour separation is 0.005~mag. A y-axis
      elongated central structure can be observed at HST resolution (see also
      Fig.~\ref{fig:center}).}
    \label{fig:hstcol}
  \end{center}
\end{figure*}

The two HST frames were then PSF-crossconvolved and divided to obtain
a $V - I$ colour image (Fig.~\ref{fig:hstcol}). It reveals a central gradient (V-I
increases towards the center) with a $y$-axis elongated structure 
centered at \mbox{(-0\farcs02, 0\farcs01)} (see dashed line on
Fig.~\ref{fig:center}). The isochromes have slightly higher ellipticities than
the isophotes ($\varepsilon=0.5$ at 2\farcs5 for $V-I$, vs.  0.4 for the $V$ frame).


\begin{figure*}
  \centering \resizebox{\textwidth}{!}{\includegraphics{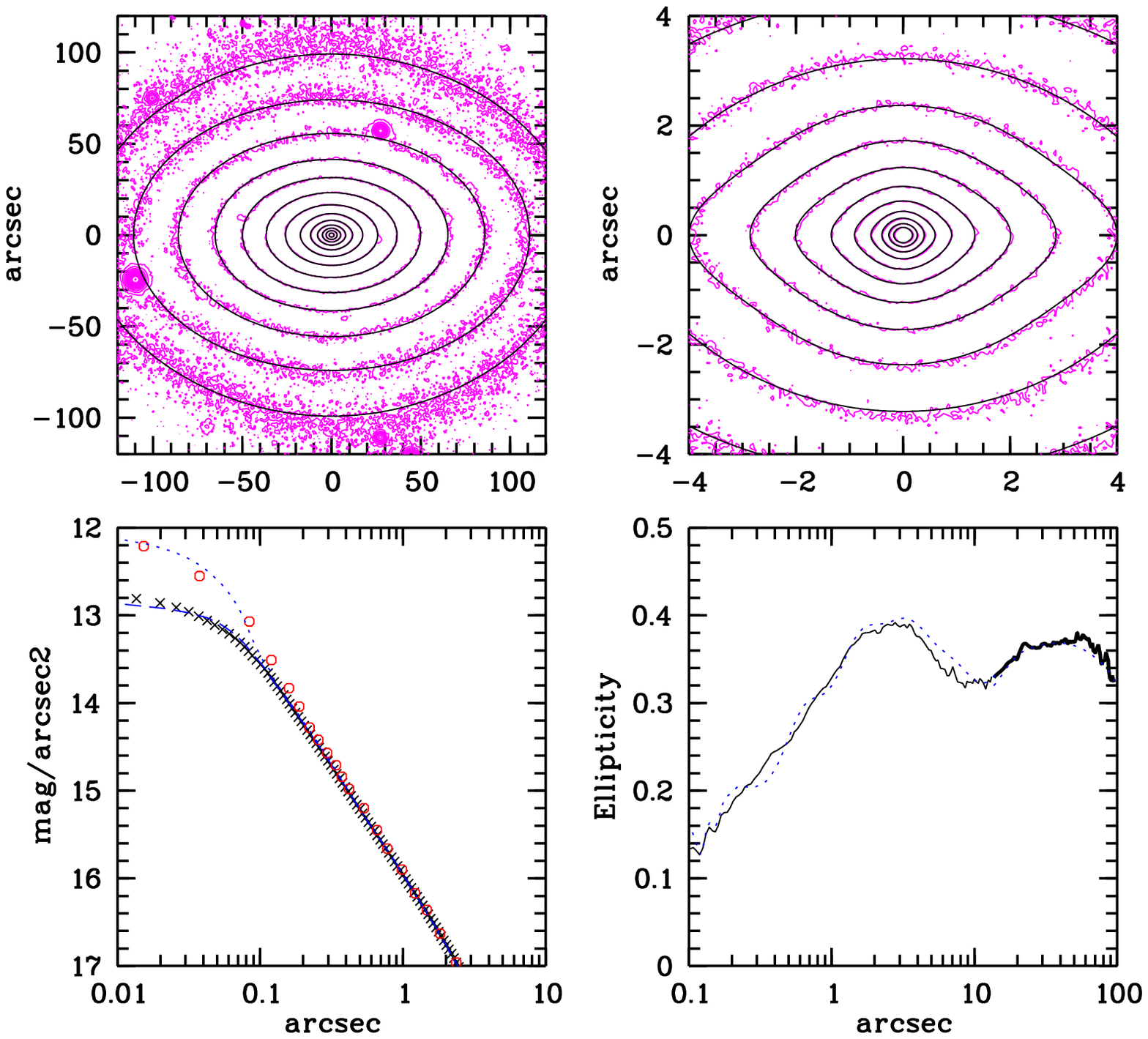}}
  \caption{Top panels: Multi Gaussian Expansion fit (thick contours)
    of NGC~4621 superimposed on the $V$ band isophotes (thin contours). Top
    left~: OHP $V$-band photometry. Top right~: HST/WFPC2 F555W band image
    (isophote step of 0.4 mag/arcsec$^2$). Notice the nuclear disc in the HST
    data. Bottom left panel: NGC~4621 light profiles along
    $r^2=x^2/a^2+y^2/b^2$, where $a$ and $b$ are the semi major and minor axes
    of fitted ellipse respectively. Crosses correspond to the WFPC2 image,
    circles to the deconvolved WFPC1 data from Byun \etal~(\cite{paper2}).
    The dashed and dotted curves correspond to the convolved and deconvolved
    MGE models respectively.  Bottom right panel: ellipticity profile, with
    Michard's (solid bold line) and WFPC2 (solid hairline) data, along with
    the MGE model (dotted line).}
\label{fig:mgefit}
\end{figure*}

\section{Models}
\label{sec:DF}

In this Section, we present the methods used to model the photometry and
kinematics of the central region of NGC~4621. The available data suggest that
the very central region (100~pc) of NGC~4621 slightly departs from
axisymmetry. The CRC seems off-centered in both the \oasis\ and \stis\ data.
However, the off-centering is only $\sim 4$~pc. At the resolution of the BSG94
data, this is obviously not resolved.  Moreover, the position angle measured
in the OHP photometry does not vary more than 2 degrees within 15\arcmin.
Axisymmetry is therefore a reasonable approximation for a first modelling. 
Axisymmetric, two-integral models then have the advantage to be semi-analytical. 
We first derived simple Jeans models to roughly
constrain the input parameters (inclination, mass-to-light ratio). We then
used the Hunter \& Qian (\cite{hun}) formalism to compute the two-integral
distribution function $f$ of NGC~4621, as a function of energy ($E$) and of the
vertical component of angular momentum ($L_z$).  The present modelling is only
intended to provide a first view at the dynamics of NGC~4621, so we 
decided not to include a central dark component: this issue will be examined
in detail in a forthcoming paper.

\subsection{MGE}
\newcommand{\proj}{}
\begin{table}
\begin{center}
  \begin{tabular}{rrrr}
    \hline
    & $\proj{I} (L_{\sun}.pc^{-2})$
    & $\proj{\sigma}(\arcsec)$ & $\proj{q}$ \\ 
    \hline \hline 
    1 & 4.502 \ten{5} &    0.040 & 0.860  \\ 
    2 &  7.713 \ten{4} &    0.112 & 0.610  \\ 
    3 &  4.792 \ten{4} &    0.201 & 0.941  \\ 
    4 &  1.161 \ten{4} &    0.438 & {\bf 0.344}  \\ 
    5 &  1.475 \ten{4} &    0.516 & 0.919  \\ 
    6 &  4.638 \ten{3} &    1.036 & {\bf 0.325}  \\ 
    7 &  6.317 \ten{3} &    1.280 & 0.872  \\ 
    8 &  3.357 \ten{3} &    2.486 & {\bf 0.275}\\ 
    9 &  2.700 \ten{3} &    3.211 & 0.658  \\ 
    10 &  1.326 \ten{3} &    5.698 & 0.817  \\ 
    11 &  6.313 \ten{2} &    6.926 & {\bf 0.377}  \\ 
    12 &  6.417 \ten{2} &   12.468 & 0.639  \\ 
    13 &  3.295 \ten{2} &   25.674 & 0.627  \\ 
    14 &  8.208 \ten{1} &   57.091 & 0.633  \\ 
    15 &  1.517 \ten{1} &  128.782 & 1.000  \\ 
    \hline 
  \end{tabular}
\end{center} 
\caption{MGE photometric model of NGC~4621 in the $V$ band. 
  The flattened inner disc components are emphasized in bold.}
\label{tab:mge}
\end{table}

The first step in modelling NGC~4621 was to build a luminosity model which
properly reproduced the observed photometry.  We used the Multi Gaussian
Expansion (MGE) formalism proposed by Monnet \etal~(\cite{monnet}) and
Emsellem \etal~(\cite{mge}), which expresses the surface brightness as a sum
of two-dimensional Gaussians. Assuming the spatial luminosity is also a sum of
(three-dimensional) Gaussians, given the choice of viewing angles, and using
an MGE model for the PSF, we could then deconvolve and deproject the MGE model
uniquely and analytically.

The procedure was performed stepwise. We began fitting the wide field $V$-band
image. We then subtracted the outer gaussian components from the high
resolution WFPC2 image, and fitted the residual image (central 15\arcsec). In
this step, we had to exclude the innermost 0\farcs4, to avoid convergence
problems probably due to the slightly asymmetric central feature
(Fig.~\ref{fig:center}). We finally fitted the very central arcsecond.
Gathering these three parts, we thus obtained a 15 gaussian components model,
with the same center and PA, the parameters of which are given in
Table~\ref{tab:mge}. The goodness of the fit is illustrated in
Fig.~\ref{fig:mgefit}.

\subsection{Two-integral models}

We then made use of the Hunter \& Qian (\cite{hun}) formalism to build the
two-integral distribution function of the galaxy using the best fit value for
the mass-to-light ratio of $\Upsilon_V=6.6\ M_\odot/L_\odot$ found from simple
Jeans models. We used a default value of $i = 90\degr$, which produced a
marginally better fit.

\begin{figure}
  \resizebox{\columnwidth}{!}{\includegraphics{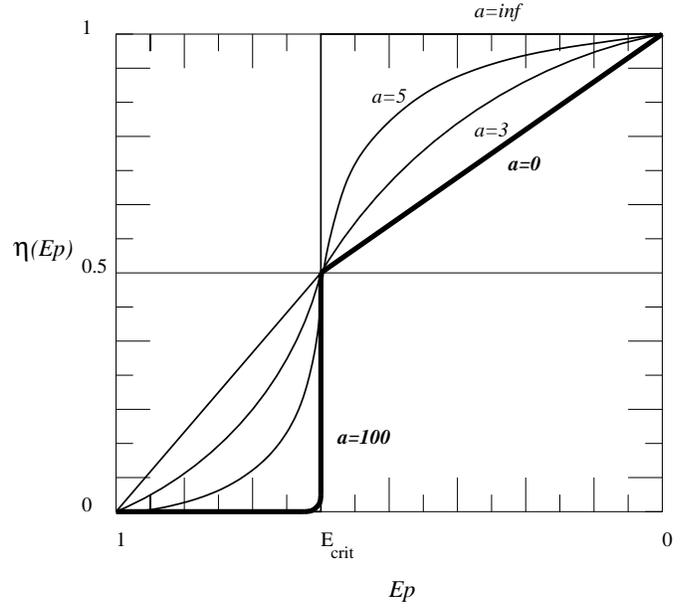}}
  \caption{Parametrization of $\eta$. $E_p$ is the normalized energy
    $E_p=E/E_{\mbox{\tiny max}}$. $E_{\mbox{\tiny crit}}$ is the energy at
    which the counter-rotation starts, \eg when $\eta<0.5$, $f_{\mbox{\tiny
        odd}}<0$ (see Eq.~\ref{eqn:param}). The bold line corresponds to the
    best-fitting model.}
  \label{fig:eta}
\end{figure}

The DF is divided into two parts, which are respectively even and odd in
$L_z$. The even part ($f_{\mbox{\tiny even}}$) is uniquely determined by the
input MGE mass model.  This involves the calculation of a path integral in the
complex plane, as described by Hunter \& Qian (\cite{hun}).  The odd part of
the distribution function ($f_{\mbox{\tiny odd}}$) is then parametrized, and
adjusted to fit the kinematical data. We chose the parametrization proposed by
van der Marel \etal~(\cite{vdm}), and modified it to account for the CRC. The
original parametrization corresponds to Eq. \ref{eqn:param}, and we
additionally allowed $\eta$ to be function of $E_p=E/E_{\mbox{\tiny max}}$
(Fig.~\ref{fig:eta}).  The analytical form of this $\eta$ function is the same
as the one in Eq. \ref{eqn:hpar}, with an additional variable change. We can
adjust the energy $E_{\mbox{\tiny crit}}$ above which the stars begin to
counter-rotate, as well as the smoothness of the transition ($a=\infty$:
abrupt transition, $a=1$: smooth transition, see Fig.~\ref{fig:eta}).

\newcommand{\arctanh}{\mbox{arctanh}}
\begin{eqnarray}
        f_{\mbox{\tiny odd}}(E,L_z)&=&(2\eta-1)\ h_{\alpha}(L_z/L_{z,\mbox{\tiny
        max}})\cdot f_{\mbox{\tiny even}}(E,L_z) \label{eqn:param}\\
        h_{\alpha}(x)&=&\left\{
        \begin{array}{ll}
                \tanh(\alpha x/2)/\tanh(\alpha/2)       & (\alpha>0)\\
                x                            & (\alpha=0)\\
                (2/\alpha)\arctanh[x\tanh(\alpha/2)]   & (\alpha<0)
        \end{array}\right.
        \label{eqn:hpar}
\end{eqnarray}

\begin{figure}
  \begin{center}
    \resizebox{\columnwidth}{!}{\includegraphics{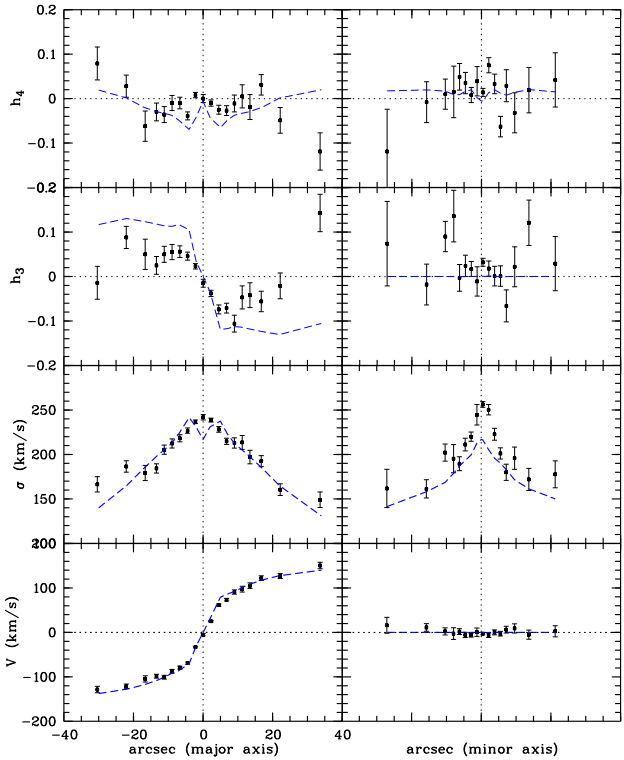}}
    \caption{Best fit two-integral DF model. Left: Major axis. Right: Minor
      axis. Inclination is 90 degrees, no black hole and $\Upsilon_V=6.06$.
      Points are the BSG94 data, and the curve is the model. The fit is
      reasonable but indicates the need for a black hole (low central
      $\sigma$) and maybe a third integral (overestimated $h_3$).}
    \label{fig:plotcompn}
  \end{center}
\end{figure}

\begin{figure}[h]
  \begin{center}
    \resizebox{\columnwidth}{!}{\includegraphics{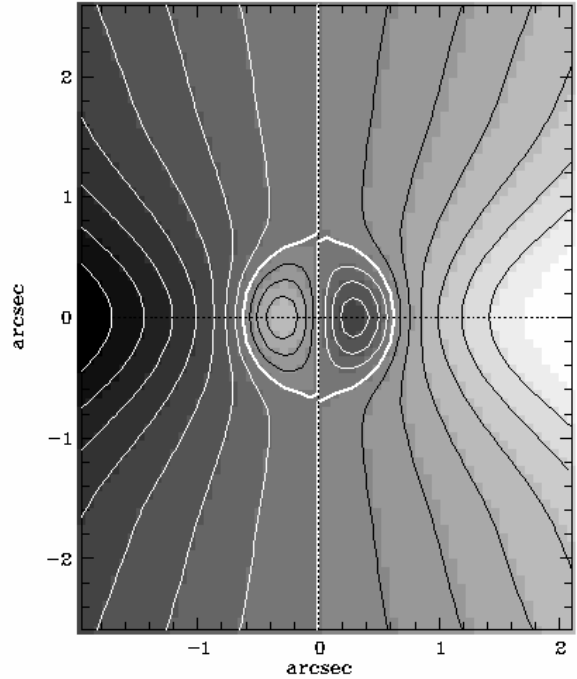}}
    \caption{Best fit two-integral DF model for the comparison with
      \oasis. Only the velocity map is shown: central dispersion values are
      too low due to the absense of a central dark mass. Grey levels range
      from -50 to +50~\kms}
    \label{fig:plotoasis}
  \end{center}
\end{figure}

The best fit model reproduces the BSG94 velocity profiles reasonably well
(Fig.~\ref{fig:plotcompn}) with values of $\alpha$ ranging from 8 outside
10\arcsec\ to -2 in the central part. The higher resolution \oasis\ 
(Figs.~\ref{fig:plotoasis} and \ref{fig:VGH}), and \stis\ 
(Fig.~\ref{fig:HRdata}) velocity measurements, revealing the counter-rotating
core, are also well fitted by this two-integral model.  The best fitting model
uses a core with a diameter of 1\farcs1 ($E_{\tiny\mbox{crit}}=0.62$), and an
abrupt transition ($a=100$ \emph{i.e.}  almost all stars having
$E_p>E_{\tiny\mbox{crit}}$ are counter-rotating, see bold line in
Fig.~\ref{fig:eta}). A rough estimate of the CRC mass can be made by selecting
stars counter-rotating in the central part. This is performed by integrating
the DF weighted by a function which is 0 for $ E < E_{\tiny\mbox{crit}}$ and
$L_z > 0$ and 1 otherwise. The total mass of NGC 4621, which is given by the
mass-to-light ratio and the deprojected MGE-model is $1.78\cdot10^{11}\ 
M_{\odot}$. The total mass of the CRC is $2.13\cdot10^{8}\ M_{\odot}$,
yielding a mass fraction of 0.12\%.

The dispersion profiles are well reproduced by the model outside the central 
few arcseconds. The central values of the dispersion predicted by the 
models are however systematically too low compared to the BSG94 observations. 
This is confirmed by the \oasis\ and \stis\ dispersion values: the model
predicts a central dispersion of $\sim 220$~\kms, to be compared with
the central observed \stis\ dispersion of $320 \pm 27$~\kms.
We were finally unable to fit the higher order moments, even at large radii.
The $h_3$ values predicted by the model
are thus systematically too high, by a factor of almost two.
This discrepancy could not be solved even by changing the parameters
of the odd part of the DF.

These two discrepancies do indicate that we need a more general dynamical
model for NGC~4621. First, we should remove the constraint imposed
by the two-integral model by allowing a third integral of motion.
There may then still be the need for an additional central dark mass
to explain the observed dispersion values. Such a model will be examined
in a forthcoming paper (Wernli et al., in preparation).

\section{Discussion and conclusion}
\label{sec:discussion}

In this paper, we report the discovery of a $\sim 60$~pc diameter
counter-rotating core using new \oasis\ integral field spectroscopy, a result
confirmed with archival \stis\ long-slit data.  A structure at this scale has
never been observed before, except in M~31, which exhibits a 20~pc co-rotating
decoupled core (Bacon \etal~\cite{m31} and references therein). Typical sizes
of kinematically decoupled cores are at the order of 1 kpc (See \eg the sample
of Carollo \etal~\cite{marc}). The computed mass fraction of 0.12\%
($2.1\cdot10^{8}\ M_{\odot}$) is on the low side, as compared for instance to
the CRC of IC~1459, which is 0.5\% for a core size of 3 kpc (Cappellari
\etal~\cite{ic1459}), but for which the estimation is of higher accuracy. As a
comparison, the black hole mass from the $\sigma-M_\bullet$ relation for
IC~1459 is $1.1\cdot10^{9}\ M_{\odot}$ (Merritt \& Ferrarese \cite{bhmasses2})
and $2.13\cdot10^{8}\ M_{\odot}$ for NGC~4621 (Merritt \& Ferrarese
\cite{bhmasses}). The mass fraction estimate provided here in the case of
NGC~4621 can only be indicative: the DF separation is inaccurate and
difficult, and to get a precise result we would need an analysis in integral
space.


Both individual \oasis\ and \stis\ data sets suggest that the CRC of NGC~4621
is slightly off-centered with respect to the center of the outer isophotes.
This offset seems consistent with the weak asymmetry detected in the WFPC2/HST
$V$ and $I$ frames and the $V-I$ colour map (Fig.~\ref{fig:center}). The
central structure elongated along the minor axis detected in the $V-I$ colour
map represents an increase of $V-I=0.03$ which could either be due to an
intrinsic stellar population gradient, or to dust. In the latter case, this
would correspond to $A_V=0.06$ (assuming $R_V=3.1$ and $R_I=1.86$ from models
of the Galaxy, Rieke \& Lebofsky \cite{dust}). Note that we do not detect any
high frequency structure in the individual WFPC2 $V$ and $I$ frames.

The two-integral models reasonably reproduce the observed velocity profiles in
the outer parts, as well as in the region of the CRC (\oasis\ and \stis).
These axisymmetric models do obviously not take into account the observed
off-centering. The predicted central value of the velocity dispersion is
significantly too low by 50~\kms. We were also unable to simultaneously fit
higher order Gauss-Hermite moments: \eg the predicted $h_3$ is systematically
too high (see Fig.~\ref{fig:plotcompn}). This demonstrates that we need more
general dynamical models in order to properly fit the observed kinematics.  We
are therefore in the process of constructing three-integral Schwarzschild
models, with the possibility to include a central dark mass. This may solve
the discrepancies mentioned above.

If one assumes a merger-scenario for the CRC's origin, the off-centering, if
confirmed, could be either due to the fact that the merging process is still
on-going, or may be the result of a stable mode (see \eg M~31 in Bacon \etal~
\cite{m31}). The short dynamical timescale at the radius of the CRC (1~Myr)
seems to favour the latter hypothesis. Further discussions regarding this
issue must however wait for additional spectroscopy. A detailed study at high
spatial resolution of the stellar populations in the central arcsecond would
certainly help in this context. Existing high resolution \stis\ spectra at
bluer wavelengths can be used for this purpose.

Stellar kinematics at HST resolution is available for only a handful of
early-type galaxies. M~31 is one of the rare examples where it is possible to
have sufficient spatial resolution to measure kinematical features at parsec
scales, such as its 20~pc nucleus (Bacon et al. \cite{m31} and references
therein). The OASIS data presented in this paper demonstrate that it is
possible to study substructures with a characteristic size of 60~pc in
galaxies at the distance of the Virgo cluster. The presence of the CRC in
NGC~4621 raises new questions about the dynamical status of the centers of
early-type galaxies. What is the fraction of early-type galaxies that have
such substructures? Two-dimensional spectroscopy at high spatial resolution is
clearly needed to simultaneously study the dynamical and chemical contents of
these cores, and to eventually understand their origin.


\begin{acknowledgements}
 We would like to thank Raymond Michard and Ralf Bender for the data they provided,
Tim de Zeeuw for a careful reading of the manuscript and his suggestions, 
and the referee Dr M. Carollo for her thorough and constructive report.
\end{acknowledgements}


\bibliographystyle{aabib99}

\end{document}